# Temperature dependence of lower critical field in stripe ordered La$_{1.6-x}$Nd$_{0.4}$Sr$_x$CuO$_4$ superconductors

L. Xie[1], Y. W. Yin[2,*], X. G. Li[2,3]

[1]Department of Physical and Chemical Sciences, North China University of Technology, Beijing 100144, P. R. China

[2]Hefei National Laboratory for Physical Sciences at Microscale, Department of Physics, University of Science and Technology of China, Hefei 230026, P. R. China

[3] School of Physics and Materials Science, Anhui University, Hefei 230601, China

**Abstract:**

We report detailed measurements of the temperature dependence of the lower critical field $H_{c1}$ in the stripe ordered La$_{1.6-x}$Nd$_{0.4}$Sr$_x$CuO$_4$ ($x$ = 0.10 and 0.15) superconductors. It is found that $H_{c1}(T)$ of the samples increase with decreasing temperature and show an negative curvature below $T_c$. The penetration depth $\lambda(T)$ estimated from $H_{c1}(T)$ data satisfies a linear temperature dependence behavior below $T_c/2$, which is consistent with that in a superconductor with $d$-wave symmetry of the energy gap and the upward trend in $H_{c1}(T)$ may be due to the cooperation of the hole and election pockets arising from the reconstruction of the Fermi surface, which contributes to illustrating the Fermi surface reconstruction in La$_{1.6-x}$Nd$_{0.4}$Sr$_x$CuO$_4$.

---

[*]Author to whom correspondence should be addressed. Email: yyw@ustc.edu.cn

**Introduction**

The recent observation of quantum oscillations in under-doped high-$T_c$ superconductors, combined with their negative Hall coefficient at low temperature, reveal that the Fermi surface of hole-doped cuprates includes a small electron pocket [1-3]. Considering the large hole Fermi surface characteristic of the superconductors in over-doped regime, this strongly suggests that the Fermi surface undergoes a reconstruction caused by the onset of some order which breaks the translational symmetry. The most possible order is the stripe phase order, a charge/spin modulation observed clearly in materials like $La_{1.6-x}Nd_{0.4}Sr_xCuO_4$ (LNSCO), in which the onset of stripe order coincides with major changes in transport properties, providing strong evidence that stripe order is indeed the cause of Fermi surface reconstruction [4-8]. The combination of positively and negatively charged current carriers may provide a key to understanding cuprate superconductors [9].

Another crucial issue in understanding the superconducting mechanism is the pairing symmetry of the superconducting gap and the nature of the low energy excitations. Lower critical field $H_{c1}$, or equivalently, magnetic penetration depth $\lambda$ is a fundamental probe of the nature of the pairing symmetry, since the appearance of gap nodes strongly modifies the $T$ dependence of the superfluid density and thereby the penetration depth [10]. Compared with other probing technologies, it is advantageous that $H_{c1}(T)$ measurement probe relatively large distances ($\lambda \sim 100$ nm) and are far less sensitive to sample surface quality [11]. Although the superconducting gap of LNSCO has been studied by the scanning tunneling microscopy/spectroscopy [12], little data has been focused on $H_{c1}(T)$ which is also important for probing the pairing symmetry.

In the present paper, we present detailed magnetization measurements on the lower critical field of the stripe ordered LNSCO superconducting single crystals. And the pairing symmetry of LNSCO is analyzed according to the penetration depth $\lambda(T)$.

**Experiments**

The $La_{1.6-x}Nd_{0.4}Sr_xCuO_4$ (LNSCO, $x = 0.10$ and 0.15) single crystals used here are grown by the traveling solvent floating zone technique as reported previously [13-14]. The grown crystals are carefully cut into rectangles along the main crystalline axes. The typical sample size for the magnetization measurements is $2.0\times1.0\times0.7$ mm$^3$ with the shortest edge along the $c$ axis. The crystals are annealed in air at 800 ºC for 48 $h$ to attain proper oxygen content and achieve homogeneity before measurements. The quality of our samples has also been characterized by X-ray diffraction patterns which show that the crystals are of a pure phase and the full width at half maximum of peak (004) is about 0.15º, shown in Fig. 1, indicating a good quality of the crystals. All the magnetization measurements were carried out by a Vibrating Sample Magnetometer (VSM) (Quantum Design).

**Results and discussion:**

Figure 2 shows the susceptibility of the $La_{1.6-x}Nd_{0.4}Sr_xCuO_4$ samples $x = 0.10$ and 0.15. The $T_c$ of the samples $x = 0.10$ and 0.15 defined as the onset of the sharp drop in susceptibility are about 7.6 K and 14.3 K with the transition width of $\Delta T_c = 5.6$ K and 11.3 K, respectively, which are due to the two-dimensional superconductivity caused by the stripe phase order [15].

The isothermal magnetization curves $M(H)$ were taken at different temperatures in a zero-field-cool (ZFC) mode with initial temperature up to 35 K above $T_c$. The field was changed in the No-overshoot mode with 1 Oe as a step. In Fig. 3(a), we display the typical initial $M(H)$ curves at selected temperatures. It can be seen that, at low fields all curves clearly show a common linear dependence of the magnetization on field. For a strict treatment, we can fit data points between 0 and 20 Oe at $T = 2$ K by a linear law for each set of curves of samples $x = 0.10$ (not shown here) and 0.15, respectively. These fitted linear lines describe the common Meissner shielding effects ("Meissner line") at low fields, as shown by the red solid lines in Fig. 3.

As we know, the accurate determination of $H_{c1}$ by magnetization measurements may suffer from demagnetization effect. $H_{c1}$ can be deduced from the first penetration field $H_{c1}^*$, assuming that the magnetization $M = -H_{c1}$ when the first vortex enters into the sample. Thus $H$ has been rescaled to $H_{eff} = H - NM$ and $H_{c1} = H_{c1}^* / (1-N)$, where $N$ is the demagnetization factor and $H$ the external field. It has been shown by Brandt that a bar with a rectangle cross-section, the effective demagnetization factor $N = 1 - \tanh\sqrt{0.36 b/a}$, where $a$ and $b$ are the width and the thickness of the sample [16]. Using this equation we estimated the demagnetizion factor $N \sim 0.54$ for our sample of dimensions $a = 2.0$ mm and $b = 0.7$ mm. We transform the magnetic moment $m$ in emu unit into $M$ in Oe unit with the relation $1 emu/cm^3 = 4\pi$Oe and the $H$ into $H_{eff} = H - NM$, as displayed in Fig. 3(b). The slope of the fitted line is -0.999, very close to -1 ($M = -H_{c1}$), demonstrating the reliability of our data.

We determine the value of $H_{c1}$ by examining the point of departure from the Meissner line on the initial slope of the magnetization curves. Fig. 4 shows the initial magnetization curves after subtracting the "Meissner line" as a function of $T$ for sample $x = 0.15$. The results for sample $x = 0.10$ are not shown here. By using a criterion of $\Delta M = 3 \times 10^{-4}$ emu, which is close to the resolution limit of our measurement system, we determined the lower critical field $H_{c1}$. The extracted nominal $H_{c1}$ as a function of $T$ for these two samples are presented in Fig. 5(a), and both show a negative curvature and an upward trend with decreasing temperature. The upward trend will not change by using different criterion for $\Delta M$, such as $10^{-4}$ emu or 0.5 Oe as used in reference [11], for the determination of $H_{c1}$.

In order to investigate the temperature dependence of the penetration depth $\lambda$ in the low-temperature region, we use the expression $H_{c1} = \frac{\Phi_0}{4\pi\lambda^2} \ln\kappa$, where $\Phi_0 = hc/2e = 2 \times 10^{-7}$ G cm$^2$ is the flux quantum, and $\kappa$ is the Ginzburg-Landau parameter. Although we have not measured the value $\kappa$ for LNSCO here, the curves of $\lambda(T)$ can be estimated, as shown in Fig. 5(b) where the $C = \sqrt{\frac{4\pi}{\Phi_0 \ln\kappa}}$ is a constant. This is because $\kappa$ is a weak temperature-dependent function and the $\ln\kappa$ can be taken as a constant [17]. It is remarkable that the dependence of $\lambda$ on temperature for $x = 0.10$ below 4 K and for $x = 0.15$ below 6.5 K ($T_c/2$ for samples) is almost linear, which indicates that the derivation $\Delta\lambda(T)$ of the penetration depth from its zero-temperature value $\Delta\lambda = \lambda(T) - \lambda(0)$ is a linear relation. This linear dependence contradicts the prediction for an isotropic $s$-wave superconductivity, for which the $\Delta\lambda$ is given by $\frac{\Delta\lambda(T)}{\lambda(0)} \approx 3.3 \left(\frac{T_c}{T}\right)^{1/2} \exp\left(-\frac{\Delta}{k_B T_c}\frac{T_c}{T}\right)$, where $\Delta$ is the energy gap [18]. A $T^n$ variation of normalized $H_{c1}(T)$, e.g., $\Delta\lambda(T) \propto T^n$ with n = 1 and 2, at low temperatures is the character of *d*-wave state with line nodes in superconducting gap function [19, 20]. Moerover, the slope of $d\lambda(T)/dT$ at 0 K is finite in LNSCO $x$ = 0.10 and 0.15 according to this linear relation. In cuprate superconductors, the finite slope of $d\lambda(T)/dT$ at 0 K has been ascribed to the *d*-wave symmetry of the energy gap [11, 22] based on the consideration as follows. In a pure *d*-wave superconductor, the energy gap along the node directions ($k_x = \pm k_y$) vanishes and the spectrum $N_s(E)$ of quasiparticle excitations in the superconducting phase is gapless, resulting in the linear dependence of $N_s(E)$ on $E$ at low excitation energies. A finite temperature will excite certain quasiparticles leading to a linear dropping of the superfluid density $\rho_s$ with temperature. Therefore, the above results imply the *d*-wave symmetry for LNSCO $x$ = 0.10 and 0.15, which is consistent with that obtained by the scanning tunneling microscopy [12].

Now let's return to $H_{c1}(T)$. As can be seen from Fig. 5(a), $H_{c1}$ is linear in temperature below $0.6T_c$, indicating line nodes in superconducting energy gap, consistent with *d*-wave pairing [23]. However, for LNSCO $x$ = 0.15, there is an upward trend in low temperature. Usually, the lower critical field $H_{c1}(T)$ shows a downward trend in cuprates, such as $La_{1.86}Sr_{0.14}CuO_4$ [24], $YBaCuO_{6+x}$ [23]. An upward curvature of $H_{c1}(T)$ near $T_c$ was found in $Pr_{2-x}Ce_xCuO_4$ and explained successfully using a weakly coupled two-band model implying the existence of two kinds of charge carriers [25]. Moreover, recently, the upward trend was found in FeAs based superconductors such as $Ba_{0.60}K_{0.40}Fe_2As_2$ [26] and $SmFeAsO_{0.9}F_{0.1}$ [11]. This conflicts with the single band gap description of the mean field theory, and hence has been used as evidence of two energy gaps like $MgB_2$ or $FeTe_{0.6}Se_{0.4}$ superconductors [27, 28]. Through the density functional study, Subedi *et.al*. showed that the band structure of FeSe and FeTe consist cyndirical electron Fermi surface at the zone corner and two concentric cynderical hole surface a the zone center [29], indicating that the superconductivity in $FeTe_{0.6}Se_{0.4}$ results from two bands, and the upward curvature of $H_{c1}$ in $FeTe_{0.6}Se_{0.4}$ is caused by the coexistence of both electrons and holes [28].

Recently, several experimental studys have confirmed that the Fermi-surface reconstruction could occur in various cuprates, such as $YBa_2Cu_3O_y$ [30], $La_{1.6-x}Nd_{0.4}Sr_xCuO_4$ [31], $La_{1.8-x}Eu_{0.2}Sr_xCuO_4$ [32], $HgBa_2CuO_{4+x}$ [5], and it appears to be common for all hole-doped cuprates. For example, Doiron-Leyraud *et al*. have given a direct evidence through quantum oscillations experiments that in the underdoped $YBa_2Cu_3O_y$ there are small electron [33] and hole [34] pockets at low temperature. The Fermi surface reconstruction is probably caused by the onset of some order which breaks the translational symmetry and the stripe phase order is the most possible one. The transport properties, such as Hall effect, Seebeck effect, Nernst effect, change sharply at the onset of the stripe order and turn to be negative, providing strong evidence that stripe order is indeed the cause of Fermi surface reconstruction and the electron pocket will have an influence on the properties [4-8].

The Fermi surface of the LNSCO, a famous stripe phase superconductor, also undergoes a reconstruction at low temperature, resulting in the coexistence of the electron and hole pockets in the Fermi surface. In a two band model, the superfluid density proportional to $H_{c1}(T)$ can be expressed as $\rho_s(T) = \rho_{s,1}(T) + \rho_{s,2}(T)$ [25]. Due to the finite gap in the quasiparticle excitations in the first band, $\rho_{s,1}(T) = \rho_{s,1}(0)(1 - ae^{(-\Delta_1/k_B T)})$, where $\Delta_1$ is the energy gap of the first band, $a$ is a constant. There are gap nodes in the second band, therefore $\rho_{s,2}(T)$ should behave similarly as in a pure $d$-wave superconductor and show a linear $T$ dependence in low temperatures due to the low energy linear density of states, $\rho_{s,2}(T) = \rho_{s,2}(0)(1 - \frac{T}{T_c})$. Assuming the transition temperature for the two bands are $T_{c1}^0$, $T_{c2}^0$ and $T_{c1}^0 < T_{c2}^0$. Just below $T_c$, $\rho_s(T)$ is mainly contributed from the second band. However, when $T$ drops below $T_{c1}^0$, the intrinsic superconducting correlation of the first band will appear in addition to the induced one, and the contribution to $\rho_s(T)$ from this band will rise rapidly with decreasing temperature. Consequently, a clear upturn will show up at low temperatures. The two band model have been successfully used to explain the upward trend of $H_c(T)$ in $Pr_{2-x}Ce_xCuO_4$ [25] and iron-based superconductors [26-29]. Therefore, the upward curvature of $H_{c1}$ in LNSCO may be also due to the cooperation of the hole and electron pockets and the accurate mechanism needs deep investigations in future. Anyway, the experimental results obtained here may contribute to illustrating the Fermi surface reconstruction in LNSCO.

**Conclusion**

In conclusion, we have measured the *M-H* curves of LNSCO $x = 0.10$ and $0.15$ samples and obtained the lower critical field $H_{c1}(T)$. A striking upward trend of $H_{c1}(T)$ has been observed at low temperature for $x = 0.15$, which may be dictated by both the hole and electrons pockets. The penetration depth $\lambda(T)$ satisfies a linear temperature dependence implying the *d*-wave symmetry of the energy gap for LNSCO.

**Acknowledgement**:

This work was supported by the National Natural Science Foundation of China, the National Basic Research Program of China (Contract Nos. 2011CBA00102, 2012CB922003, and 2015CB921201), and Beijing Municipal Education Commission (KM201510009012).

**Figure captions:**

**Fig. 1.** X-ray diffraction patterns for samples LNSCO $x = 0.10$ and 0.15 single crystals.

**Fig. 2**. Temperature dependent susceptibility of LNSCO $x = 0.10$ and 0.15 single crystals. The inset shows the schematic diagram of the samples. The arrows indicate the $T_c$ for the samples.

**Fig. 3.** (a) Magnetization hysteresis loops of LNSCO $x = 0.15$ at various temperatures; (b) Magnetic moment *M* in Oe as a function of the effective magnetic field $H_{eff} = H - NM$ considering the demagnetization factor *N*. The two red solid lines in (a) and (b) are the "Meissner line" showing the linearity of these curves at low fields, as discussed in the text.

**Fig. 4**. The initial part of the magnetization curves *M(H)* of LNSCO $x = 0.15$ at various temperatures obtained after subtracting "Meissner lines" from the raw data. The dashed line defines the criterion of $\Delta M = 3 \times 10^{-4}$ emu used to determined $H_{c1}$.

**Fig. 5**. (a) Temperature dependence of $H_{c1}$ for LNSCO $x = 0.10$ and 0.15. The red solid lines show the linear relation; (b) The estimated values of $\lambda(T)$ with a constant

$C = \sqrt{\dfrac{4\pi}{\Phi_0 \ln \kappa}}$ . The solid lines indicate the clear linear $T$ dependence of $\lambda(T)C$.